%% file: example_paper.tex
\documentclass{article}

\usepackage{microtype}
\usepackage{graphicx}
\usepackage{subfigure}
\usepackage{booktabs} 

\usepackage{hyperref}

\usepackage[preprint]{icml2024}

\usepackage{amsmath}
\usepackage{amssymb}
\usepackage{mathtools}
\usepackage{amsthm}

\usepackage[capitalize,noabbrev]{cleveref}

\theoremstyle{plain}

\theoremstyle{definition}

\theoremstyle{remark}

\usepackage[textsize=tiny]{todonotes}
\usepackage{pifont}
\newcommand{\xmark}{\ding{55}}%
\input{mathcommand}

\begin{document}
\twocolumn[
\icmltitle{TrustMol: Trustworthy Inverse Molecular Design via Alignment with Molecular Dynamics}

\begin{icmlauthorlist}
\icmlauthor{Kevin Tirta Wijaya}{mpi}
\icmlauthor{Navid Ansari}{mpi}
\icmlauthor{Hans-Peter Seidel}{mpi}
\icmlauthor{Vahid Babaei}{mpi}
\end{icmlauthorlist}

\icmlaffiliation{mpi}{Max Planck Institute for Informatics, Saarbrücken, Germany}

\icmlcorrespondingauthorpreprint{ }{\{kwijaya, nansari, hpseidel, vbabaei\}@mpi-inf.mpg.de}

\icmlkeywords{Machine Learning, ICML}

\vskip 0.3in
]

\printAffiliationsAndNoticePreprint{} 
\begin{abstract}
Data-driven generation of molecules with desired properties, also known as inverse molecular design (IMD), has attracted significant attention in recent years.
Despite the significant progress in the accuracy and diversity of solutions, existing IMD methods lag behind in terms of \textit{trustworthiness}.
The root issue is that the design process of these methods is increasingly more implicit and indirect, and this process is also isolated from the \textit{native forward process} (NFP), the ground-truth function that models the molecular dynamics.
Following this insight, we propose TrustMol, an IMD method built to be trustworthy. 
For this purpose, TrustMol relies on a set of technical novelties including a new variational autoencoder network. Moreover, we propose a latent-property pairs acquisition method to effectively navigate the complexities of molecular latent optimization, a process that seems intuitive yet challenging due to the high-frequency and discontinuous nature of molecule space. TrustMol also integrates uncertainty-awareness into molecular latent optimization. 
These lead to improvements in both explainability and reliability of the IMD process.
We validate the trustworthiness of TrustMol through a wide range of experiments.

\end{abstract}
\section{Introduction}
The discovery of new molecules with desired properties has been pivotal in driving technological advancements throughout recent history.
For instance, the synthesis of ammonia, which is essential for producing inorganic fertilizers, has been vital to keep nearly half of the world's population away from hunger \citep{smil1999detonator_ammonia_worldhunger}.
Similarly, recent molecular discoveries have improved the performance of grid-scale \citep{chen2018manganese_battery} and high-voltage \citep{zheng2020cyclic_battery} batteries, and inhibit tumour drivers \citep{yan2014discovery_drugdiscovery} as well as the viral replication of SARS-CoV-2 \cite{riva2020discovery_drugdiscovery}.
Despite its importance, traditional methods for molecular discovery are often slow and complex, relying on experts to propose, synthesize, and evaluate newly-designed molecules in an iterative manner.
An emerging solution to accelerate the molecular discovery process is through data-driven, typically deep-learning based, inverse molecular design (IMD) \citep{gebauer2022inverse_inverseMolecularDesign,gebauer2019symmetry_3dGenerative, luo20213d_3dGenerative, hoogeboom2022equivariant_edm, xu2023geometric_geoldm}. 
In contrast to the more commonplace forward simulation of molecular properties, in IMD molecules are designed given a target property.  
Although the ultimate goal of IMD is to invert the \textit{native forward process} (NFP), i.e., the ground-truth function that maps a molecule to its corresponding properties according to the molecular dynamics, devising an inverse NFP is extremely challenging, and the common approach is to resort to a surrogate model built using data sampled from NFP.

If we aspire for data-driven IMD methods to be widely adopted by their eventual users, such as chemists, we must build \textit{trustworthiness} into them.
Trustworthiness encompasses several dimensions as outlined by the \citet{eu2019_trustworthyAI} of the European Commission. In this work we limit the scope of trustworthiness to \textit{explainability} and \textit{reliability}.
We adapt the definitions of explainability and reliability for surrogate-based IMD, where explainability requires that both the NFP surrogation and the inversion method to be interpretable. 
Reliability, on the other hand, involves (1) accurate generation of molecules with desired properties, (2) internal mechanism for quality evaluation of the generated molecules, and (3) alignment of said mechanism with the NFP-based evaluation method. 
Unfortunately, current state-of-the-art data-driven IMD methods, dominated by diffusion models \citep{hoogeboom2022equivariant_edm, xu2023geometric_geoldm}, fall short in both trustworthy measures.
It is difficult to interpret the quality of the NFP surrogation, i.e., how well the surrogate model captures the underlying principles of the NFP, solely from the training metrics of diffusion models because of their unpredictable behavior \cite{github_edm_issue} and lack of physical meaning. 
Moreover, diffusion models lack any internal mechanism for evaluating the quality of the generated molecules.

Instead of further advances with diffusion models, we advocate a promising alternative to improve trustworthiness of surrogate-based IMD: direct optimization of neural surrogate models. 
This involves training a neural surrogate model to predict the properties of molecular designs, and then using an optimization to iteratively improve a randomly-initialized molecular design.
Such a direct optimization offers better explainability compared with diffusion models because we can easily interpret the quality of the NFP surrogation from the training error metrics.
Furthermore, during the inversion of the surrogate, the error between the desired properties and the properties predicted by the neural surrogate model, i.e., the \textit{surrogate error}, can be used as an immediate and internal mechanism to evaluate the quality of the generated molecules.
However, reliable surrogation of a molecular NFP, and inverting this surrogate is a formidable challenge, primarily due to the high-frequency and discontinuous characteristics of molecular dynamics, which are often adversarial for neural networks \citep{rahaman2019spectralbias, xu2019frequencybias}.
Our proposed method, TrustMol, realizes the above vision through a set of technical novelties. 
TrustMol sets forward an inverse framework for molecular design that directly optimizes a neural surrogate model in a learned latent space taking into account the uncertainty of the predictions. 
The contributions of TrustMol, which we will expand in Section \ref{sec:method}, can be summarized as follows.

\textbf{NFP-aware neural surrogate-based optimization.}
TrustMol utilizes an NFP-aware optimization approach which is well-suited for improving the interpretability of the surrogation of the NFP and assessing the quality of the generated molecules.
This approach not only allows the IMD process to be more explainable, but also more reliable compared with state-of-the-art diffusion models as it is more aligned with the molecular dynamics.\\
\textbf{Optimization in a latent space.}
To address the challenge of the NFP surrogation, TrustMol performs the optimization in a learned molecular latent space.
TrustMol employs the SGP-VAE, a novel variational autoencoder (VAE) that incorporates both string and three-dimensional (3D) graph molecular representations, alongside molecular properties, for constructing this latent space.
As a result, similar latents are more likely to exhibit similar properties, allowing the neural surrogate model to fit the mapping from the latent space to the property space.
While optimization in the molecular latent space has been explored before\citep{gomez2018automatic_latentOpt, eckmann2022limo}, our novelty lies in the molecular latent-space construction that incorporates both molecular string and 3D graphs.\\
\textbf{VAE reconstruction error compensation.}
Due to the non-zero reconstruction error of the VAE, we propose to \textit{calibrate} the dataset used to train the latent-to-property neural surrogate model.
TrustMol introduces a simple latent-property pairs acquisition method that is pivotal in enabling accurate prediction of molecular properties directly from their latent representations, which ultimately improves the accuracy of the IMD process.\\
\textbf{Uncertainty-aware optimization.}
By incorporating epistemic uncertainty quantification into the optimization process, TrustMol successfully improves the \textit{alignment} of the surrogate-based and NFP-based evaluation mechanisms, resulting in a reduction of gap between the surrogate error and the NFP error.\\
\textbf{Trustworthy inverse molecular design evaluation.}
As an additional contribution, TrustMol proposes a rigorous NFP-based evaluation that can be used to benchmark various IMD methods in a trustworthy manner.
The proposed evaluation consists of measuring the NFP-based accuracy of IMD methods in generating molecules with both in-distribution and out-of-distribution target properties.

The source code for TrustMol will be made available upon publication. 
We also provide a tool (Appendix \ref{app:interactive_tool}) which is interactive thanks to precomputation in order to demonstrate the potential of the IMD process of TrustMol. We encourage the readers to experiment with this tool at: \url{https://repo012424.streamlit.app/}.

\section{Related Work}
\label{sec:related_work}
Computational methods for molecular design can be traced back to as early as 1969 when \citet{hansch1969quantitative_qsar} proposed the Quantitative Structure-Activity Relationships (QSAR) method for analyzing the biochemical structure-activity problems based on a regression model.
Since then, numerous methods have been proposed, ranging from improvements of the QSAR method \citep{kubinyi19933d_3dqsar, vedani1998quasi_3dqsar} to genetic algorithms \citep{glen1995genetic_geneticAlgorithm, venkatasubramanian1995evolutionary_geneticAlgorithm, sundaram1998parametric_geneticAlgorithm}.
In recent years, deep learning has started to influence the field.
For instance, \citet{gebauer2019symmetry_3dGenerative}, \citet{luo20213d_3dGenerative} and \citet{luo2022autoregressive_3dGenerative} introduce models that can generate molecules in an autoregressive manner.
E-NF \citep{garcia2021n_enf} proposes an equivariant normalizing flows method that integrate an equivariant graph neural network as a differential equation to obtain an invertible function.
EDM \citep{hoogeboom2022equivariant_edm} utilizes EGNN \citep{satorras2021n_egnn} as the backbone of an equivariant diffusion model.
GeoLDM \citep{xu2023geometric_geoldm} takes the diffusion approach further by performing diffusion in a latent space that is constructed with a variational autoencoder (VAE, \citet{kingma2013auto_vae}).

To generate molecules with desired properties, \citet{gebauer2019symmetry_3dGenerative} and \citet{luo20213d_3dGenerative} fine-tune the pretrained G-SchNet and G-SphereNet models on a subset of the dataset in which the molecules exhibit the desired properties.
While straightforward, this strategy comes at a cost of the controllability of the IMD process.
For example, it is challenging to use this strategy to generate a molecule that exhibits a HOMO-LUMO gap of exactly x eV.
\citet{hoogeboom2022equivariant_edm} and \citet{xu2023geometric_geoldm}, on the other hand, achieve conditional generation of molecules by using the property values of the molecules as additional training inputs to the denoising network of the EDM and GeoLDM models, respectively.

Closest to our proposed approach are the methods proposed by \citet{gomez2018automatic_latentOpt} and \citet{eckmann2022limo} (LIMO), which invert a surrogate model to directly optimize a randomly-initialized molecular latent.
A core distinction is that TrustMol constructs the latent space with considerations to molecular 3D geometry, molecular string, and molecular properties.
Moreover, TrustMol takes into account the uncertainty of the surrogate model's predictions during the molecular latent optimization.

\section{Trustworthy Inverse Molecular Design} 
\label{sec:method}

In this section, we discuss the components of TrustMol that enable a trustworthy inverse molecular design (IMD).
Additionally, in Section~\ref{sec:eval_method}, we propose a rigorous native forward process (NFP)-based evaluation that can be used as a reliable benchmark aligned with the real-world evaluation for future research in this area.

\subsection{NFP-aware Neural Surrogate-based Optimization}
A trustworthy IMD method should be explainable and reliable.
In the context of molecular design, an important aspect of explainability is the ability to measure the quality of the surrogation, i.e., how well the surrogate model $\hat{f}$ capture the underlying molecular dynamics behavior of the NFP $f$.
Apart from the training stage, a reliable IMD method should accurately generate molecules and give a measure of the quality of the generated molecular designs preferably quickly. 
An example of such a quality measure is the \textit{surrogate error}, which is the error between the desired properties and the properties of the molecules as predicted by the surrogate model. 
The surrogate error should be a reliable proxy for the \textit{NFP error}, which is the true measure of quality calculated with the NFP (more on this topic in Section~\ref{sec:surrogate}). 

Current state-of-the-art of inverse molecular design (IMD) is dominated by deep generative models, particularly diffusion models \citep{hoogeboom2022equivariant_edm, xu2023geometric_geoldm}.
However, these models often fall short in explainability and reliability of the IMD process.
For instance, it is difficult to interpret the quality of the surrogation in diffusion models \citep{github_edm_issue}.
Moreover, the surrogates in diffusion models, i.e., the denoising networks, do not possess an internal mechanism to calculate the surrogate error.
As a result, we must resort to the resource-intensive NFP to evaluate all of the generated molecular designs to find the best candidates.

One promising solution to address these issues is through direct molecular design optimization with neural surrogate model \citep{gomez2018automatic_latentOpt, eckmann2022limo}.
A neural network $\hat{f}$ can be trained to predict molecular properties $\vp_x \in \sR^d$ of a molecular design $\vx$, where $d$ represents the number of molecular properties that we are interested in.
Given desired molecular properties $\vq \in \sR^d$, a pretrained $\hat{f}$ can then be used to iteratively improve a randomly-initialized molecular design to get the optimal molecular design $\vx^*$ through gradient descent and backpropagation by minimizing the surrogate error, 
\begin{equation}
    \vx^* = \argmin_{\vx}\frac{1}{d}\sum_{i=1}^d |\hat{f}(\vx)_i - q_i|,
\end{equation}
where $f(\vx)_i = p_{x, i}$ and $q_i$ are the \textit{i}-th components of the vectors $\vp_{\hat{x}}$ and $\vq$, respectively.
This \textit{direct optimization} approach offers improvements in both explainability and reliability of the IMD process.
For example, the training error metrics of $\hat{f}$ provide an intuitive understanding of how well the neural surrogate model captures the underlying principles of the NFP.
If $\hat{f}$ predicts mean absolute errors of 0 eV and 1 eV for Highest Occupied Molecular Orbital (HOMO) and Lowest Unoccupied Molecular Orbital (LUMO) values, respectively, we can intuitively conclude that the surrogation quality for HOMO is higher than LUMO. 
In terms of reliability, the surrogate error (during inverse design) can serve as an internal mechanism to evaluate the quality of the generated molecular designs, which is not possible to do with existing diffusion models.
This is particularly beneficial for selecting high-quality candidates for more resource-intensive experimental validations, thereby avoiding unnecessary resource allocation for sub-optimal molecular designs.

A neural surrogate-based direct optimization approach for inverse molecular design might be intuitive but not straightforward. 
The major challenge is the high-frequency and discontinuous nature of the mapping from the molecule space to the property space, an issue we later study in Section~\ref{sec:highfrequencymap}.
Since neural networks often struggle to fit high-frequency functions \citep{xu2019frequencybias, rahaman2019spectralbias} and are inherently continuous, neural surrogate-based optimization in the molecular space often yields molecules with low surrogate but high NFP error, or even completely invalid molecules.
Therefore, in the following sections, we introduce several novel mechanisms that are employed by TrustMol to enable a trustworthy inverse molecular design.

\subsection{Optimization in the Latent Space}
To address the high-frequency and discontinuity issues, we propose molecular design optimization in a latent space learned by a variational autoencodoer (VAE), rather than directly in the molecule space.
This approach enables us to impose additional constraints during VAE training that promote a smoother mapping from the latent to the property space, where similar latents correlate with similar molecular properties.
Different from previous molecular optimization in latent space methods \citep{gomez2018automatic_latentOpt, eckmann2022limo}, we construct the latent space with consideration to three features: molecular string, molecular 3D geometry, and molecular properties.

Concretely, our novel SELFIES-Graph-Property (SGP) VAE network incorporates SELFIES strings, three-dimensional (3D) molecular graphs, and molecular properties during the construction of the latent space.
We use SELFIES \citep{krenn2020self_selfies} as our molecular representation to ensure that every latent representation can be translated into a valid molecule, thus resolving the discontinuity issues.
However, similarities between molecular strings do not always correspond to similar properties.
For instance, the SMILES strings  `C\textbackslash 1=C\textbackslash CC/1' and `C\textbackslash 1=C(\textbackslash C)C/1' exhibit low Levenshtein Distance but a significant difference in dipole moment (0.708D as per \citet{nist_dipole_database}).
Since solely relying on molecular strings is not sufficient to solve the high-frequency nature of the mapping to the property space, we augment the VAE training with two additional tasks: reconstructing 3D molecular graphs and predicting properties directly from the latents.
These tasks improve the smoothness of the mapping from latent to property, leveraging the observed correlation between structural similarity in 3D space and similarity in property space \citep{martin2002structurally_similarMolecules}.

As illustrated in Figure~\ref{fig:inverse_design}A, our SGP-VAE architecture features an encoder $\Phi_{\text{enc}}$ with a multi-layer perceptron (MLP) to process SELFIES strings and an equivariant graph neural network (EGNN, \citep{satorras2021n_egnn}) branch for encoding 3D molecular graphs.
Meanwhile, the decoder $\Phi_{\text{dec}}$ consists of three MLPs, one for each of the reconstruction and prediction tasks.
To train the SGP-VAE, given a 3D graph $\vx_{\text{graph}}$, a SELFIES string $\vx_{\text{selfies}}$, one encoding-decoding forward pass consists of the following,
\begin{equation}
    \vz = \Phi_{\text{enc}}(\vx_{\text{graph}}, \vx_{\text{selfies}}),
\end{equation}
\begin{equation}
    (\hat{\vx}_{\text{graph}}, \hat{\vx}_{\text{selfies}}, \hat{p}) = \Phi_{\text{dec}}(\vz),
\end{equation}
and the loss function is calculated as follows,
\begin{multline}
    \mathcal{L} = |p_{\vx} - \hat{p}_{\vx}| + ||\vx_{\text{graph}} - \hat{\vx}_{\text{graph}}||_2^2 + \\ \text{CE}(\vx_{\text{selfies}}, \hat{\vx}_{\text{selfies}}) + \text{KL}(\vz || \mathcal{N}(0, 1)),
\end{multline}
where CE and KL are cross-entropy and KL-divergence \citep{kullback1951information_kldiv} loss functions, respectively.

With the learned latent space, the neural surrogate model $\hat{f}$ is trained to predict molecular properties directly from the latents.
Consequently, the IMD is effectively a minimization of the surrogate error to obtain the optimal latent, 

\begin{equation}
 \vz^* = \argmin_{\vz} \frac{1}{d}\sum_{i=1}^d |\hat{f}(\vz)_i - q_i|,
\end{equation}
where $\vz$ is randomly-initialized from $\mathcal{N}(0, \mI)$.
After the optimization, the pretrained SGP-VAE decodes the latent back into a SELFIES string, i.e., $\vx^*_{\text{selfies}} = \Phi_{\text{dec}}(\vz^*)$.

\subsection{VAE Reconstruction Error Compensation}
\citet{eckmann2022limo} observed that directly predicting molecular properties from latents as above is challenging since the model can not generalize about what makes a molecule bind well. They instead trained the surrogate model to predict properties from the molecular string.
Here, we show that an appropriate training strategy can enable accurate predictions of properties directly from latents.

The issue stems from the fact that even after extensive training on a large dataset such as QM9 ($\sD_{\text{QM9}}$, \citep{ramakrishnan2014quantum_qm9}), the VAE still demonstrates a non-zero reconstruction error of the SELFIES strings.
This means that the decoder cannot reconstruct the original SELFIES strings perfectly, $\hat{\vx}_{\text{selfies}} \neq \Phi_{\text{dec}}(\Phi_{\text{enc}}(\vx_{\text{selfies}}))$.
This reconstruction error might be negligible in other domains such as computer vision, but it is problematic for inverse molecular design due to the high-frequency and discontinuous nature of molecule to property mapping. 
Consider, for instance, a molecular design $\vx_{\text{a}}$ that is transformed into a latent vector $\vz_{\text{a}}$ with the encoder $\Phi_{\text{enc}}$.
The decoder $\Phi_{\text{dec}}$ then transforms $\vz_{\text{a}}$ back into the molecule space.
However, due to the reconstruction error, $\vz_{\text{a}}$ is transformed into $\vx_{\text{b}}$ instead of $\vx_{\text{a}}$.
The issue is that, due to the high-frequency nature of the NFP $f$, it is possible that $||f(\vx_{\text{a}}) - f(\vx_{\text{b}})||_2 >> 0$.
As a result, a naive training strategy for the neural surrogate model such as to minimize $||\hat{f}(\Phi_{\text{enc}}(\vx)) - f(\vx)||_2^2 \hspace{0.5em} \forall \vx \in \sD_{\text{QM9}}$ could result in an inaccurate surrogate.
In other words, given a target property $q = f(\vx_{\text{a}})$, it is possible that, after optimization, the surrogate error is low but the NFP error is high because $\vz^a$ is mis-decoded into $\vx_{\text{b}}$. 

To mitigate this issue, we propose a simple method to create latent-property pairs by sampling the latent space and calculating the corresponding molecular properties with the NFP.
Utilizing a trained decoder $\Phi_{\text{dec}}$ alongside a conformer generator $h$ (RDKit, \citep{rdkit}) and the NFP $f$ (Psi4 \citep{smith2020psi4}), we generate the new dataset $\sD_{\text{new}}$ of latent-property pairs according to the following steps,
\begin{equation}
    \sZ_{\text{new}} = \{\vz_{\text{new, i}} \hspace{0.3em} | \hspace{0.3em} \vz_{\text{new,i}} \sim \mathcal{N}(0, 1)\}_{i = 1}^N,
\end{equation}
\begin{gather}
    \sP_{\text{new}} = \{f(h(\Phi_{\text{dec}}(\vz_{\text{new, i}}))) \hspace{0.3em} | \hspace{0.3em} \forall \vz_{\text{new, i}} \in \sZ_{\text{new}}\},\\
     \sD_{\text{new}} = \{(\vz_{\text{new, i}}, p_{\text{new, i}}) \hspace{0.3em} | \hspace{0.3em} \forall \vz_{\text{new, i}} \in \sZ_{\text{new}} \text{ and } \forall p_{\text{new, i}} \in \sP_{\text{new}}\}.
    \label{eq:propFromLatent}
\end{gather}
This simple latent-property pairs acquisition method proves to be highly effective in improving the inverse design performance of TrustMol, as we discuss in Section \ref{sec:ablation}.

\subsection{Uncertainty-aware Latent Optimization}
\label{sec:surrogate}
The latent-property pairs acquisition enables a relatively accurate surrogate model. 
However, such a surrogate model yields high \textit{NFP-surrogate error gap} in our experiments.
An NFP-surrogate error gap is the difference between the NFP error and the surrogate error.
A high NFP-surrogate error gap indicates that the surrogate model is not \textit{aligned} with the NFP, rendering the internal quality evaluation method unreliable.
We hypothesize that this is due to the fact that a surrogate model can become unreliable when predicting (effectively extrapolating) molecular designs that are significantly different from designs available in the training dataset. 
Such an extrapolation could yield an inaccurate prediction of the properties.
Therefore, during the inversion it is reasonable to direct the optimization toward molecular designs for which the surrogate model is deemed to be accurate.
For this purpose, we incorporate the \textit{epistemic} uncertainty into optimization \citep{ansari2022autoinverse}.
Epistemic uncertainty can be seen as the measure of data sparsity within a region that is available during training.
Therefore, minimizing epistemic uncertainty is equivalent to guiding the optimization toward molecular latents that are novel, but not completely different from latents that are available in the training dataset.

The epistemic uncertainty can be quantified by measuring the predictive disagreement between accurate but diverse neural surrogate models \cite{lakshminarayanan2017simple_deepensembles}.
We define the ensemble neural surrogate model $\hat{f}$ as a set of \textit{n} similar MLPs but with different activation layers to fit the NFP, i.e., $\hat{f} = \{\hat{f}_i \hspace{0.3em} | \hspace{0.3em} \hat{f}_i: \vz \mapsto p_{\vz}\}_{i = 1}^n$.
The epistemic uncertainty (UA) of a latent vector $\vz$ can then be defined as,
\begin{gather}
    \text{UA}(\vz) = \frac{1}{N} \sum_{i = 1}^{N} (\hat{f}_i(\vz))^2 - (\hat{f}_{\text{avg}}(\vz))^2,\\
    \hat{f}_{\text{avg}}(\vz) = \frac{1}{N} \sum_{i = 1}^{N} \hat{f}_i(\vz),
\end{gather}

\begin{figure*}[ht]
    \centering
    \includegraphics[width=0.73\textwidth]{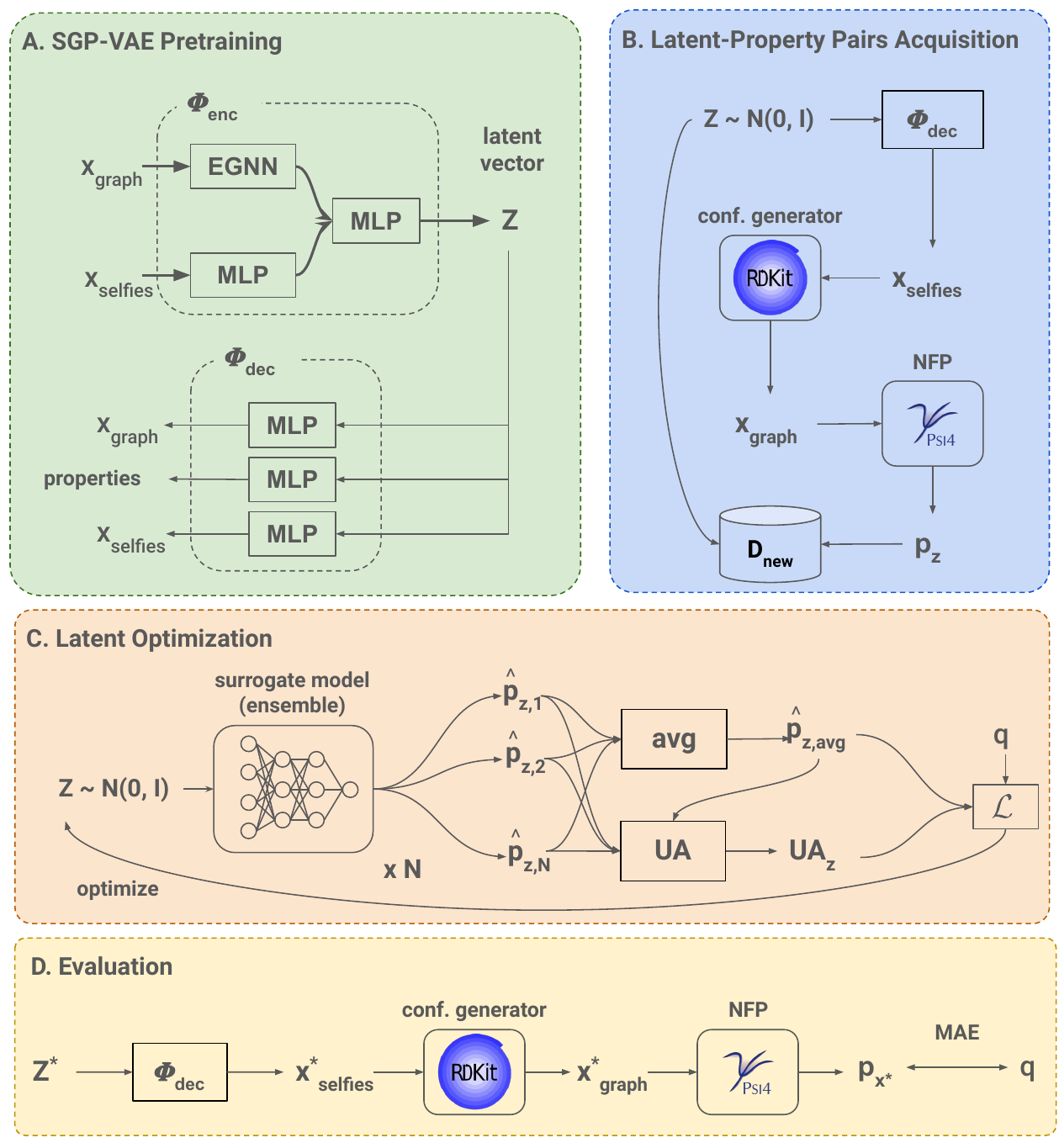}
    \caption{The overall pipeline of TrustMol. (A) shows the the training of SGP-VAE, in which $\Phi_{\text{dec}}$ is trained to reconstruct $\vx_{\text{graph}}$ and $\vx_{\text{selfies}}$, and predict the properties $\vq$ from the latent vector $\vz$. (B) is the latent-property pairs acquisition method to create a new dataset for training the surrogate model. (C) shows the latent optimization process to generate $\vz$ from $\vq$. (D) is the post-processing and evaluation, in which the optimal latent $\vz^*$ is decoded back into molecular string $\vx^*_{\text{selfies}}$, and evaluated based on the NFP.}
    \label{fig:inverse_design}
\end{figure*}

Having computed an ensemble surrogate that predicts the property and its uncertainty, the final (uncertainty-aware) IMD process of TrustMol (Figure~\ref{fig:inverse_design}~{C}) obtains a molecular latent $\vz^*$ by minimizing the following objective
\begin{equation}
    \vz^* = \argmin_{\vz}\left(\frac{1}{d} \sum_{i = 1}^d|\hat{f}_{\text{avg}}(\vz)_i - q_i| + \text{UA}(\vz)\right),
    \label{eq:final_objective}
\end{equation}
where $q_i \in \vq$ is the value of the $i$-th property that we are interested in.

\subsection{Trustworthy Inverse Molecular Design Evaluation}
\label{sec:eval_method}
Prior studies on inverse molecular design often assess the quality of the generated molecules by employing a neural surrogate model.
This model is typically trained on a different data split from the one used for training the generative models \citep{hoogeboom2022equivariant_edm, xu2023geometric_geoldm}.
Although this approach is computationally efficient, it has a crucial limitation: the surrogate models often fail to accurately model the actual NFP.
Consequently, this leads to evaluations that may not accurately represent the actual inverse molecular design performance of the methods.
Moreover, existing studies often evaluate the target property values from the distribution of the training dataset, which means no out-of-distribution property values is considered.

To ensure that the evaluation method accurately reflects the inverse design performance, we choose to calculate the properties of the generated molecules using Psi4 \citep{smith2020psi4}, an ab initio computational chemistry package for performing molecular dynamics modelling based on Density Functional Theory (DFT) calculations.
Since the inputs of Psi4 should be a 3D conformation of the molecule, we use RDKit \citep{rdkit} to generate the conformations of the SELFIES strings generated by TrustMol and LIMO.
We define our \textit{target} property values as a set of $n = 2000$ evenly-spaced values within a specified range $[a,b]$.
The range is determined based on the distribution of property values observed in the initial dataset $\sD_{\text{QM9}}$.
Specifically, we set the ranges to $[-10, 0]$ for HOMO, $[-4, 2]$ for LUMO, and $[0, 4]$ for dipole moment.
These ranges cover both property values present in, and absent from, the training dataset.
We define the evaluation criterion as the mean absolute error between the target and the DFT (NFP) calculated property values of the generated molecules.
Each generative method has a budget of 10 tries to generate a molecule for each target property value, and we retain only the molecule exhibiting the lowest absolute error.

\begin{table*}[ht]
    \caption{Experimental results for single-objective inverse molecular design (HOMO, LUMO, or Dipole Moment). We report the mean and standard deviation over three runs. The unit of measurements is electronvolt (eV) for HOMO and LUMO, and Debeye (D) for Dipole Moment. Note that the mean absolute error for each property is calculated only over molecules that are stable, i.e., not violating the octet rule, therefore methods with less than 100\% stability rate may have advantages.}
    \centering
    \begin{tabular}{lcccccc}
        \toprule
        \multicolumn{1}{c}{Model} & \multicolumn{6}{c}{Property} \\
        \cmidrule(lr){2-7}
         & \multicolumn{2}{c}{HOMO} & \multicolumn{2}{c}{LUMO} & \multicolumn{2}{c}{Dipole Moment} \\
         \cmidrule(lr){2-3}
         \cmidrule(lr){4-5}
         \cmidrule(lr){6-7}
         & MAE (eV) & Stable (\%) & MAE (eV) & Stable (\%) & MAE (D) & Stable (\%) \\
        \cmidrule(lr){1-7}
        LIMO \citep{eckmann2022limo} & 1.23 ± 0.18 & 100 & 0.35 ± 0.14 & 100 & 0.59 ± 0.08 & 100\\
        \small{LIMO on z} \citep{eckmann2022limo} & 1.31 ± 0.21 & 100 & 0.49 ± 0.17 & 100 & 0.82 ± 0.12 & 100\\
        GeoLDM \citep{xu2023geometric_geoldm} & 1.16 ± 0.03 & 83.1 & 0.39 ± 0.02 & 90.6 & 0.56 ± 0.03 & 100\\
        TrustMol (ours)& \textbf{0.95 ±0.06} & 100 & \textbf{0.25 ± 0.01} & 100 & \textbf{0.40 ± 0.02} & 100\\
        \bottomrule
    \end{tabular}
    \label{tab:overall}
\end{table*}

\section{Experimental Results}
In this section, we present the experimental results for TrustMol, comparing it with two state-of-the-art inverse molecular design methods: diffusion-based GeoLDM \citep{xu2023geometric_geoldm} and direct optimization-based LIMO \citep{eckmann2022limo}. 
This includes a comparison with LIMO on z \citep{eckmann2022limo}, a variant of LIMO with its surrogate model trained to predict properties directly from the latent vectors, which is the closest to our method.
We use HOMO, LUMO, and dipole moment as the target properties of the IMD methods. We elaborate further on experiments along with the implementation details in Appendix~\ref{app:details}.

\begin{table}[ht]
    \caption{Experimental results for novelty (Nov.), uniqueness (Uni.), and latency. Novelty and uniqueness results are the average over generating 6000 molecules with different single-objective targets (HOMO, LUMO, or Dipole Moment). We set the batch size to 2000 for the batch latency experiment.}
    \centering
    \begin{tabular}{lcccc}
        \toprule
         \multicolumn{1}{c}{Model} & Nov. & Uni. & \multicolumn{2}{c}{Latency (s)} \\
        \cmidrule(lr){4-5}
         & (\%) & (\%) & single & batch \\
        \cmidrule(lr){1-5}
        LIMO & \textbf{87.80} & 21.30 & 4.12 & 7.80\\ 
        LIMO on z & 81.87 & 43.26 & \textbf{3.98} & \textbf{6.75} \\
        GeoLDM & 81.06 & \textbf{94.26} & 8.67 & 1617\\
        TrustMol & 87.70 & 88.0 & 7.62 & 11.53\\
        \bottomrule
    \end{tabular}
    \label{tab:additional}
\end{table}

\subsection{Main Results}
We evaluate the performance of TrustMol and other inverse molecular design (IMD) methods in a single-objective setting, in which only one molecular property is used as the target.
We employ five metrics to evaluate the methods.
The mean absolute error (MAE) is the average absolute errors between the NFP-calculated properties of the generated designs and the target properties.
This is in contrast with previous works \citep{hoogeboom2022equivariant_edm, xu2023geometric_geoldm} that utilize neural networks to predict the properties of the generated molecules.
We define \textit{stability} as the percentage of generated molecular designs adhering to the octet rule.
We use \textit{novelty} and \textit{uniqueness} to measure the diversity of the generated molecular designs, with novelty representing the number of designs not present in the QM9 dataset, and uniqueness representing the number of unique designs generated.
We measure \textit{latency} in two ways: \textit{single}, the time to generate one molecular design individually, and \textit{batch}, the total time to generate multiple designs in parallel.

As shown in Table \ref{tab:overall}, TrustMol outperforms all methods by a substantial margin in all three property categories.
These results demonstrate that improving explainability through a neural surrogate-based latent optimization approach does not compromise inverse molecular design accuracy.

For molecular stability, GeoLDM frequently produces unstable molecules, as evidenced by its stability rates.
In contrast, LIMO and TrustMol consistently achieve a 100\% success rate in generating stable molecules, underlining the benefits of using SELFIES \cite{krenn2020self_selfies} as the molecular design representation for a reliable IMD method. %

Table \ref{tab:additional} shows the novelty, uniqueness, and latency of the IMD methods.
All methods display high novelty, indicating the effectiveness of both denoising and property prediction networks for discovering novel molecules.
However, existing optimization-based IMD methods tend to produce identical molecules, as reflected by their uniqueness.
In contrast, TrustMol attains a high score for uniqueness that is competitive with state-of-the-art diffusion model, GeoLDM.
The high uniqueness score can be attributed to the improved surrogate model of TrustMol, which, due to the latent-property pairs acquisition, has been trained on a more diverse set of latent vectors, enabling it to navigate toward more diverse latent solutions during optimization.
Similar to other optimization-based approaches, TrustMol can generate molecules within reasonable time frame, especially when compared to GeoLDM in batch generation setup where the latency of TrustMol is two orders of magnitude smaller.

\subsection{Measuring the Alignment Between the Surrogate Model and the NFP}

For a neural surrogate-based IMD method to be considered reliable, it should demonstrate a reasonable alignment between its surrogate model and the NFP.
This alignment can be evaluated by comparing the IMD errors as predicted by the surrogate (surrogate error) and those calculated by the NFP (NFP error).
In a perfect but unlikely scenario, the gap between the NFP and surrogate errors would be zero.

Table \ref{tab:nfp_surrogate_gap} shows the NFP-surrogate error gaps of LIMO and TrustMol.
We can see that the gaps for LIMO are relatively high.
On the other hand, TrustMol achieves lower NFP-surrogate error gaps across all three property categories.
These results validate our hypothesis that incorporating epistemic uncertainty into the optimization process can effectively reduce the NFP-surrogate error gap.
We provide an additional analysis on epistemic uncertainty quantification in Appendix \ref{app:epistemic}.
Given that a smaller gap indicates an improved alignment between the surrogate model and the NFP, the predictions of the surrogate model of TrustMol can be considered more reliable than those of LIMO.

\begin{table}[t]
    \centering
    \caption{NFP-surrogate error gap comparison between LIMO and TrustMol. The NFP-surrogate error gap is defined as the absolute difference between the NFP error and the surrogate error.}
    \begin{tabular}{lccc}
    \toprule
    \multicolumn{1}{c}{Model} & \multicolumn{3}{c}{NFP-Surrogate Error Gap}  \\
    \cmidrule(lr){2-4}
          & H (eV) & L (eV) & D (D)\\
    \cmidrule(lr){1-4}
    LIMO & 1.01 ±0.07 & 0.54 ± 0.06 & 1.36 ± 0.32 \\
    TrustMol & \textbf{0.89 ± 0.13} & \textbf{0.25 ± 0.01} & \textbf{0.40 ± 0.02} \\
    \bottomrule
    \end{tabular}
    \label{tab:nfp_surrogate_gap}
\end{table}

\subsection{Verifying the High-frequency and Discontinuous Nature of the Molecule Space} \label{sec:highfrequencymap}

In earlier sections, we have discussed the high-frequency and discontinuous nature of the mapping from molecular space to property space, which has motivated us to choose molecular latents as our design representation.
To validate our design choices, we analyze the impact of minimal noise injections on various molecular design representations with respect to their corresponding molecular properties.

Table \ref{tab:perturbation} shows the MAE between properties of the original and the noise-perturbed molecular designs.
When noise from a $\mathcal{N}(0, 0.1)$ distribution is injected into a randomly-chosen atom coordinate of a 3D graph, the proportion of stable molecules drastically decreases from 100\% to 38.5\%.
Additionally, the properties of the remaining stable molecules changes significantly, as indicated by the relatively high MAE values.
The same trend can be seen when the perturbation targets atom types of the 3D graphs, in which we randomly change a single atom type into another.

Interestingly, utilizing SELFIES strings as molecular representations can improve robustness to such perturbations.
For instance, replacing a randomly-selected alphabet in a SELFIES string with another valid alphabet only reduces the stability to 60.0\%, while the MAEs between the original and perturbed molecular designs show improvements.
It is important to note that while SELFIES strings can always be translated into a stable molecule, the NFP that is used to generate the corresponding 3D conformation may not always converge due to the complexity of the molecule, which flags the molecule as unstable in our evaluation.

Finally, we can see that latent representations of molecules exhibit the greatest robustness toward perturbations.
When a $\mathcal{N}(0, 0.1)$ noise is injected into the latents, the proportion of stable molecules remains high at 67.2\%, and the MAE between the properties of the original and perturbed molecules is approximately 45\% lower in average than that observed with SELFIES strings.
These results validate our explanations regarding the high-frequency and discontinuous nature of the molecule-property mapping, and support our strategy of developing a custom latent space to smooth this mapping.

\begin{table}[t]
    \centering
    \caption{Effects of small perturbations on molecular stability and property values. We randomly add $\mathcal{N}(0, 0.1)$ noise to a single atom coordinate or a single latent's component, and randomly change one atom type or one SELFIES' alphabet. We show the MAE between the original and perturbed representation for HOMO (H), LUMO (L), and Dipole Moment (D).}
    \begin{tabular}{lcccc}
    \toprule
       Perturbation & Stable & \multicolumn{3}{c}{Mean Absolute Error} \\
       \cmidrule(lr){3-5}
       On & (\%) & H (eV)& L (eV)& D (D)\\
       \cmidrule(lr){1-5}
       Graph &  & \\
       \hspace{0.3em} - 3D coordinates & 38.5 & 1.59 & 1.79 & 0.53\\
       \hspace{0.3em} - atom types & 38.0 & 1.48 & 1.44 & 0.41 \\
       SELFIES & 60.0 & 0.86 & 1.16 & 0.47 \\
       Latents & \textbf{67.2} & \textbf{0.42} & \textbf{0.47} & \textbf{0.24} \\
   \bottomrule
    \end{tabular}
    \label{tab:perturbation}
\end{table}

\subsection{Multi-Objective Inverse Molecular Design}

While single-objective inverse molecular design (IMD) has been commonly used in previous studies \citep{hoogeboom2022equivariant_edm, xu2023geometric_geoldm}, real-world applications often are more interested in multi-objective IMD.
Therefore, we provide an analysis of multi-objective IMD performance of LIMO and TrustMol.
In this comparison, the IMD methods are tasked with generating molecular designs that simultaneously exhibit specific values of HOMO, LUMO, and dipole moment.
We set the target ranges of HOMO to [-8, -3] and of LUMO to [-3, 2] to avoid scenarios where the target HOMO value is lower than the target LUMO value.

We show the multi-objective MAE of each property in Table \ref{tab:multi}.
We also provide and visualize the hypervolume metric of the Pareto front in Appendix \ref{app:hypervolume} as an aggregate metric for multi-objective IMD.
As shown in Table \ref{tab:multi}, simultaneously optimizing for multiple properties tends to reduce the accuracy of IMD methods.
Nevertheless, TrustMol manages to minimize the deterioration of its IMD accuracy, significantly outperforming LIMO in all property categories.
The superior performance of TrustMol can be attributed to the synergy of our uncertainty-aware optimization and latent-property pairs acquisition for training the surrogate model.

\begin{table}[t]
    \caption{Experimental results for multi-objective IMD.}

    \centering
    \begin{tabular}{lccc}
        \toprule
         \multicolumn{1}{c}{Model} & \multicolumn{3}{c}{Mean Absolute Error}\\
         & H (eV) & L (eV) & D (D)\\
        \cmidrule(lr){1-4}
        LIMO & 0.85 ± 0.05 & 1.02 ± 0.05 & 1.17 ± 0.11\\ 
        TrustMol & \textbf{0.62 ± 0.03} & \textbf{0.63 ± 0.02} & \textbf{0.79 ± 0.03}\\
        \bottomrule
    \end{tabular}
    \label{tab:multi}
\end{table}

\section{Discussion}
We introduced TrustMol, a molecular latent optimization method that focuses on aligning with the native forward process (NFP) for a trustworthy inverse molecular design (IMD).
TrustMol not only demonstrates superior performance over existing IMD methods in accuracy, but also excels in explainability and reliability, marking a significant step toward a trustworthy inverse molecular design.
The effectiveness of TrustMol, however, is limited by the expressiveness of the latent space and the accuracy of the surrogate model.
Therefore, improving the latent space construction and the surrogation  is crucial for a highly performant IMD.
A promising path toward this goal is to explore the latent space further with active learning. 
A major advantage of the direct optimization approach employed by TrustMol is the ease of incorporating more constraints to the IMD process, for example, to skew the process toward molecules with low molecular mass (Appendix \ref{app:regularization}).
This flexibility presents an exciting opportunity to tune the IMD process more flexibly with the practical requirements of the end users, e.g., chemists. 

\section*{Impact Statement}
This paper presents work whose goal is to advance the field of Machine Learning. There are many potential societal consequences of our work, none which we feel must be specifically highlighted here.

\bibliography{example_paper}
\bibliographystyle{icml2024}
\nocite{rdkit}


\newpage
\appendix
\onecolumn

\section{Implementation Details}
\label{app:details}

\subsection{Dataset and Molecular Properties}
We use the QM9 dataset \citep{ramakrishnan2014quantum_qm9} as our initial training dataset $\sD_{\text{QM9}}$.
QM9 is a quantum chemistry dataset that consists of around 130K small molecules.
Each molecule is represented at atomic-level, i.e., atom types and their corresponding 3D coordinates.
The molecules contains up to 9 heavy atoms (C, N, O, F), and up to 29 atoms when including the Hydrogens.
QM9 also provides various molecular properties including dipole moment, isotropic polarizability, Highest Occupied Molecular Orbital (HOMO), Lowest Unoccupied Molecular Orbital (LUMO), thermal capacity, among others.

In our experiments, we use HOMO, LUMO, and dipole moment as the potential target properties of the inversion.
The gap between HOMO and LUMO can be used to predict the stability of a compound.
Dipole moment, on the other hand, is a measure of a molecule's polarity, which in turn can be used to predict various physical properties such as solubility in water and boiling point.

\subsection{Architecture Details for SGP-VAE}
The encoder of SGP-VAE is realized with an EGNN \citep{satorras2021n_egnn} block to process the molecular 3D graph and an MLP block to process the SELFIES string.
The EGNN block consists of 3 EGNN layers, each with 192 hidden features.
The MLP block for SELFIES is constructed with 4 \texttt{nn.Linear} layers of \{64, 128, 256, 256\} hidden features and \texttt{SILU} activation function.
We encode the features extracted by the EGNN block for 3D graph and the MLP block for SELFIES into a latent vector with another MLP block similar to the one used for SELFIES.
We set the size of latent vector to 256.
The decoders of SGP-VAE are realized with MLPs.
There are three decoders in total: one for reconstructing 3D graph with 5 \texttt{nn.Linear} layers of \{512, 512, 1024, 1024, \textbf{m} $\times$ \textbf{c}\} hidden features, one for reconstructing SELFIES string with 7 \text{nn.Linear} layers of \{512, 512, 512, 512, 512, 512, \textbf{n} $\times$ \textbf{o}\} hidden features, and one for predicting the molecular properties of a latent with 5 \texttt{nn.Linear} layers of \{512, 512, 256, 256, \textbf{d}\} hidden properties.
\begin{itemize}
    \item \textbf{m}: maximum number of atoms in a 3D graph
    \item \textbf{c}: number of atom types + 3 (for 3D coordinates)
    \item \textbf{n}: maximum number of alphabet in a SELFIES string
    \item \textbf{o}: number of alphabet types
    \item \textbf{d}: number of properties to be predicted
\end{itemize}
 
\subsection{Architecture Details for Latent-to-Property Surrogate Model}
\begin{figure}[ht]
    \centering
    \includegraphics[width=0.8\textwidth]{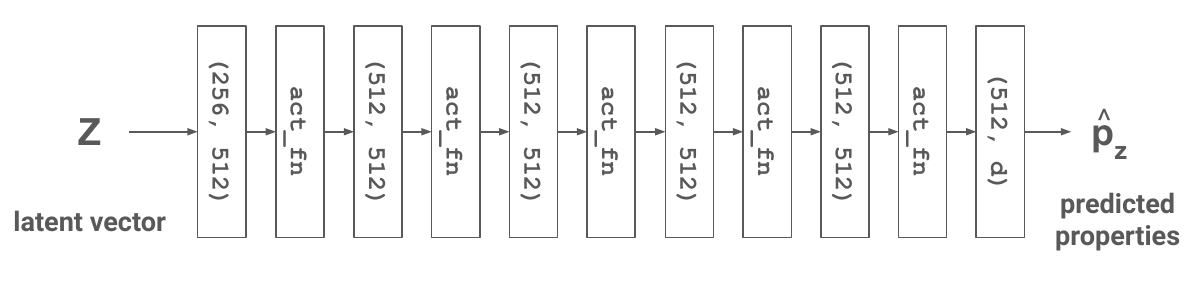}
    \caption{The architecture for the latent-to-property subnetwork. A (\texttt{x, y}) block represents an \texttt{nn.Linear} layer with an input dimensionality of \texttt{x} and an output dimensionality of \texttt{y}.}
    \label{fig:latent_to_property}
\end{figure}

The latent-to-property surrogate model is an ensemble of ten similar multi-layer perceptrons (MLPs).
Specifically, each subnetwork consists of a sequence of 6 \texttt{nn.Linear} layers with an activation function \texttt{act\_fn} in between, as illustrated in Figure \ref{fig:latent_to_property}.
To promote diversity within the ensemble surrogate model, we use a different \texttt{act\_fn} obtained from the set \{\texttt{Hardswish, LeakyReLU, ReLU, SILU, Softplus}\} for every two subnetworks .

\subsection{Training Strategy}
We implement all neural networks with PyTorch \citep{paszke2019pytorch}.
AdamW optimizer \citep{loshchilov2017decoupled_adamw} and cosine annealing learning rate scheduler \citep{loshchilov2016sgdr_cosineannealing} are used in the optimization process for all models.
We train the SGP-VAE for 50 epochs and the ensemble surrogate model for 300 epochs, with a batch size of 32.
To improve diversity of the ensemble surrogate model, at each iteration, a subnetwork $\hat{f}_i$ in the ensemble has a probability of only $q = 0.3$ to perform a gradient descent step.
This is equivalent to independently training each subnetwork for 90 epochs with different random seeds.

\subsection{RDKit and Psi4 for Native Forward Process}
We use RDKit \citep{rdkit} and Psi4 \citep{smith2020psi4} as the native forward processes, the ground truth functions that model the behavior of molecules in real-world.
RDKit is an open-source cheminformatics and machine learning software that can perform analysis on chemical structures.
We use RDKit to generate the molecular conformation, i.e., the spatial arrangement of atoms in a molecule, of the SELFIES strings generated by LIMO \citep{eckmann2022limo} and TrustMol.
Psi4 is an open-source quantum chemistry software that is capable of accurately predicting the properties of a molecular conformation using Density Functional Theory (DFT).
We use Psi4 to calculate the HOMO, LUMO, and dipole moment values of molecular conformations generated by the IMD methods.

\section{Additional Experimental Results}

\subsection{Effects of VAE Representation and Latent-Property Pairs Acquisition}
We analyze the effects of the VAE representation and latent-property pairs acquisition. 
We use HOMO as the target property of the IMD process, and calculate the mean absolute error with respect to the target property values only over molecules that are valid.

As shown in Table \ref{tab:ablation_app}, using 3D graph as both VAE and design representations leads to the worst result.
Because of the non-zero VAE reconstruction error, an optimal latent solution could be translated back into an incorrect 3D graph.
This VAE reconstruction error is exacerbated by the high-frequency and discontinuous characteristics of the molecule space, resulting in a poor IMD performance.
Simply changing the representation into SELFIES helps in improving the IMD performance, an evidence that the mapping from SELFIES to property is relatively smoother.
Note that 3D graph molecular designs are not always valid, as we have seen in Table \ref{tab:overall}, while SELFIES molecular designs are guaranteed to be always valid.

Motivated by the fact that structurally similar molecules tend to exhibit similar properties \citep{martin2002structurally_similarMolecules}, we incorporate both 3D Graph and SELFIES for constructing the latent space, but keep the molecular design representation to SELFIES only.
In this way, we get the best of both worlds, i.e., structural similarity, smoother mapping, and always-valid molecular designs.
Finally, we can eliminate the negative effects of the VAE reconstruction error by performing a latent-property pairs acquisition and creating a new dataset to train the latent-to-property surrogate model.
The utilization of this acquisition method results in a significant jump in IMD performance, enabling TrustMol to outperform other IMD methods.

\label{sec:ablation}
\begin{table}[ht]
    \caption{Ablation study on the choice of VAE representation and latent-property pairs acquisition. VAE Rep. denotes the molecular representations that are used to construct the latent space of the VAE, while Design Rep. denotes the molecular design representation that we optimize to obtain the desired molecules.}
    \centering
    \begin{tabular}{ccccc}
    \toprule
    \multicolumn{2}{c}{VAE Rep.} & Design & Latent-Property & \multicolumn{1}{c}{Mean Absolute Error} \\
    \cmidrule(lr){1-2}
    3D Graph & SELFIES & Rep. & Pairs Acquisition & HOMO (eV) \\
    \cmidrule(lr){1-5}
    \checkmark & \xmark & 3D Graph & \xmark & 1.893 \\
    \xmark & \checkmark & SELFIES & \xmark & 1.398 \\
    \checkmark & \checkmark & SELFIES & \xmark & 1.295\\
    \checkmark & \checkmark & SELFIES & \checkmark & 0.901 \\
    \bottomrule
    \end{tabular}
    \label{tab:ablation_app}
\end{table}

\subsection{Regularization of Inverse Molecular Design}
\label{app:regularization}
\begin{figure}[ht]
    \centering
    \includegraphics[width = 0.45\textwidth]{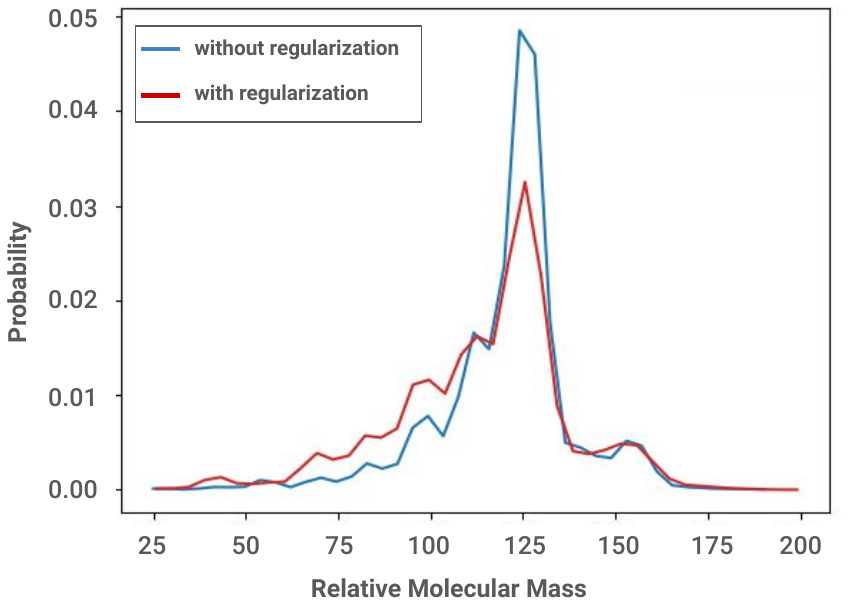}
    \caption{Additional regularizations can be easily incorporated into TrustMol. Here, we add molecular mass to the optimization objectives, penalizing molecular designs with high masses. We can see that the distribution of the generated molecular designs shifts toward molecules with lower molecular mass.}
    \label{fig:moreRegularization}
\end{figure}
Another advantage of TrustMol is the simplicity of adding more regularization into the optimization process.
For instance, suppose that we want to find molecules with not only specific HOMO, LUMO, or Dipole Moment values, but also small molecular mass.
To incorporate the new molecular mass regularization, we can train another neural network to fit the mapping from latent space to the molecular mass space, and minimizing the predicted molecular mass of the latent vectors should result in molecular designs with smaller molecular mass.
Figure \ref{fig:moreRegularization} shows the distribution of molecular designs generated with and without an additional molecular mass regularization.
As we can see, the addition of the molecular mass regularization shifts the distribution toward molecular design with smaller molecular mass.

\begin{figure}[b!]
    \centering
    \includegraphics[width = 0.45\textwidth]{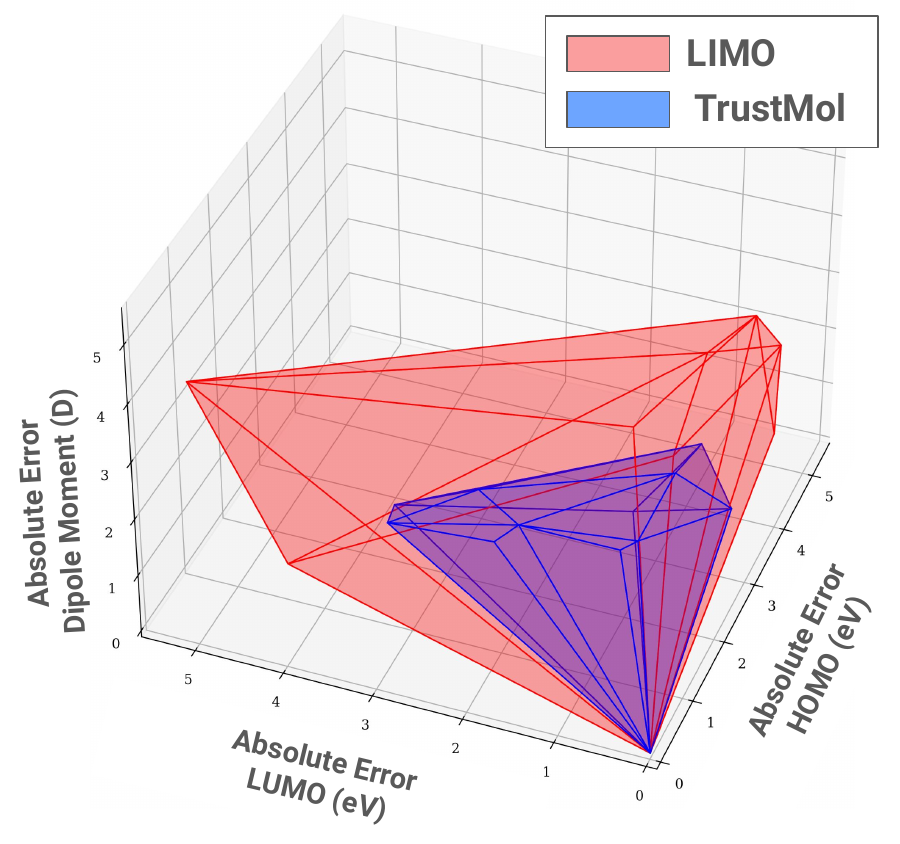}
    \caption{Visualization of the hypervolume of MAE for LIMO and TrustMol. We can clearly see the smaller space covered by the hypervolume of TrustMol.}
    \label{fig:hypervolume}
\end{figure}

\subsection{Hypervolume of Pareto Front as an Aggregate Metric for Multi-Objective IMD}
\label{app:hypervolume}

While Table \ref{tab:multi} provides a comprehensive information regarding the multi-objective IMD performance on individual property, we are also interested in assessing the multi-objective IMD as a whole with a single metric.
One candidate for such a metric is the hypervolume metric \citep{zitzler1998multiobjective_hypervolume}, which represent the size of the space covered by the Pareto frontier.

We calculate the hypervolume metrics of the mean absolute errors of LIMO and TrustMol with Pymoo \citep{pymoo}.
LIMO has a hypervolume metric of 37.117, while TrustMol has a hypervolume metric of 16.851.
Since the size of the hypervolume of MAE grows with the worst-possible errors of the IMD method, we can easily conclude that TrustMol outperforms LIMO in multi-objective IMD.
For clarity, we visualize the hypervolumes of LIMO and TrustMol in Figure \ref{fig:hypervolume}.

\subsection{The Epistemic Uncertainty for Molecular Dataset}
\label{app:epistemic}
\begin{figure}[h]
    \centering
    \includegraphics[width = 0.5\textwidth]{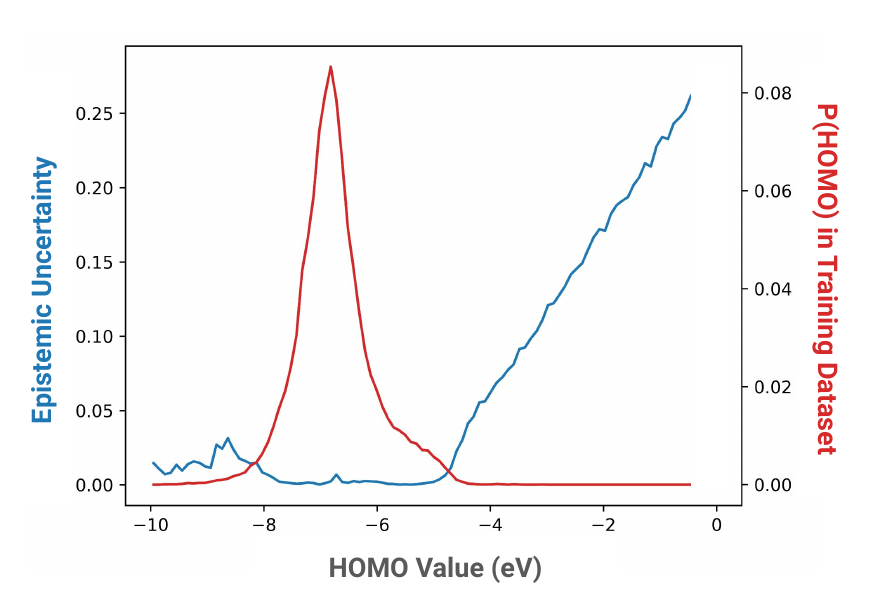}
    \caption{Plot of epistemic uncertainty values predicted by a surrogate model and the distribution of HOMO values in the training dataset.}
    \label{fig:epistemic_ua}
\end{figure}

We have discussed in Section \ref{sec:surrogate} how the epistemic uncertainty quantification helps with finding molecules that are novel but not completely different from those that are present in the training dataset.
As a supplementary analysis, we plot the epistemic uncertainty predicted by a pretrained surrogate model along with the distribution of HOMO values available in the training dataset in Figure~\ref{fig:epistemic_ua}.

As we can see in Figure~\ref{fig:epistemic_ua}, the surrogate model predicts low epistemic uncertainty when the HOMO values are densely available in the training dataset.
The reverse is true when there are few-to-no samples available in the training dataset.
Therefore, we can verify that epistemic uncertainty is a measure of data sparsity within a region that is available during training.
Adding epistemic uncertainty as an optimization objective (Equation \ref{eq:final_objective}) is then equivalent to keeping the molecular design similar to the designs found in the training dataset, which are queries of the NFP.
Since a neural network typically shows an excellent performance for in-distribution data, incorporating epistemic uncertainty ultimately leads to a better alignment between the surrogate model and the NFP.

\newpage
\section{Interactive Tool}
\label{app:interactive_tool}
\begin{figure}[ht]
    \centering
    \includegraphics[width=0.7\textwidth]{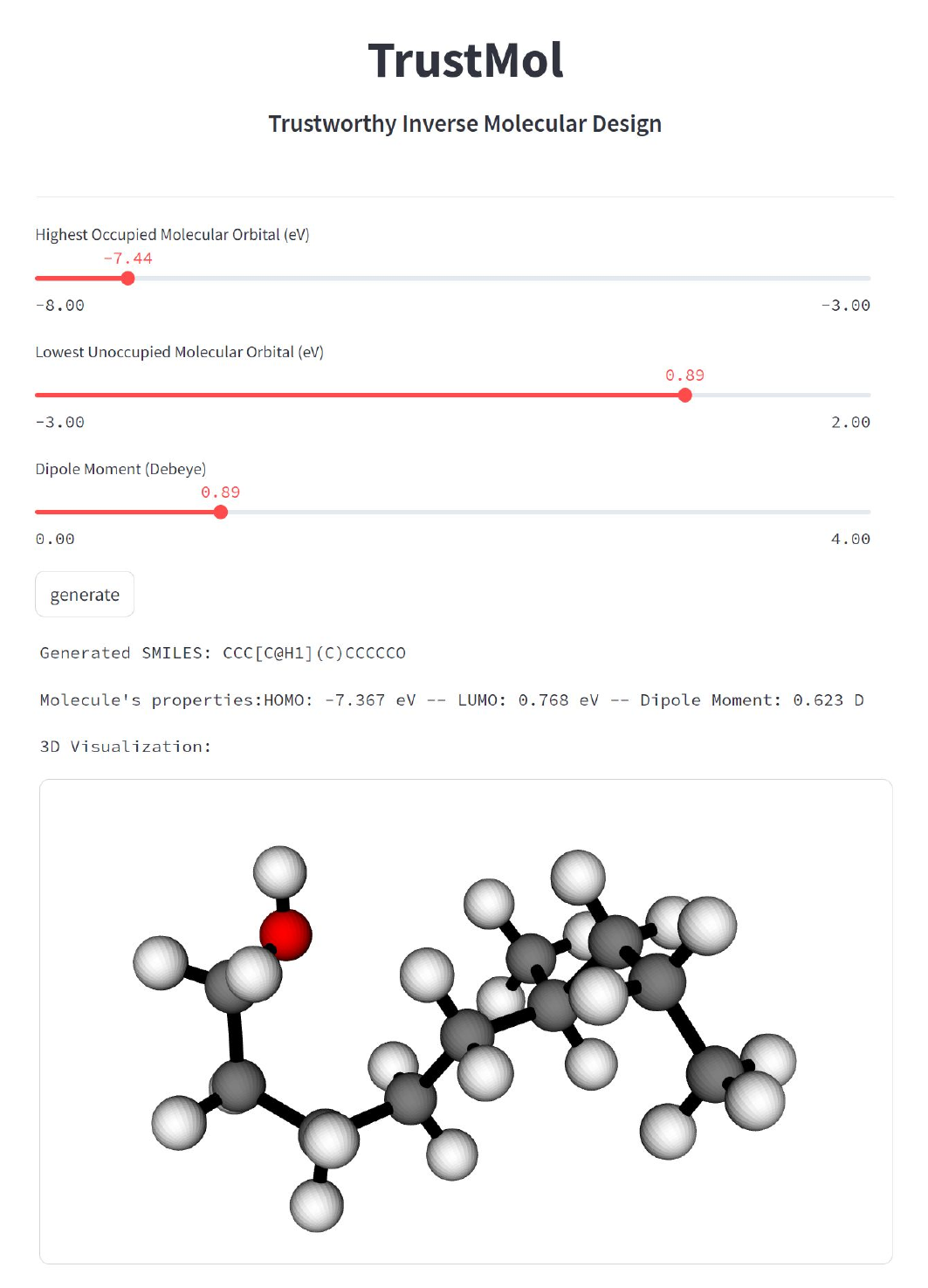}
    \caption{An interactive tool for generating molecules with TrustMol.}
    \label{fig:gui}
\end{figure}

To improve the accessibility of TrustMol for users from various background, we offer a web-based interactive tool that showcases the IMD process of TrustMol.
The user interface is shown in Figure \ref{fig:gui}, and can be accessed via \url{https://repo012424.streamlit.app/}.
We visualize the atoms in a molecule with the following colors:
hydrogen - white,
carbon - grey,
nitrogen - blue, 
oxygen - red,
fluorine - green.

Currently we use the precomputed results of the multi-objective IMD shown in Table \ref{tab:multi} since the NFP-based evaluations are too computationally demanding.
However, our ultimate objective is to fully integrate all technical components of TrustMol within this interactive tool, in line with our vision to bring inverse molecular design to its end users, i.e., chemistry and materials science practitioners.

\end{document}

%% file: mathcommand.tex
\usepackage{amsmath,amsfonts,bm}

\def\eqref#1{equation~\ref{#1}}

\def\1{\bm{1}}

\def\vp{{\bm{p}}}
\def\vq{{\bm{q}}}

\def\vx{{\bm{x}}}

\def\vz{{\bm{z}}}

\def\mI{{\bm{I}}}

\DeclareMathAlphabet{\mathsfit}{\encodingdefault}{\sfdefault}{m}{sl}
\SetMathAlphabet{\mathsfit}{bold}{\encodingdefault}{\sfdefault}{bx}{n}

\def\sD{{\mathbb{D}}}

\def\sP{{\mathbb{P}}}

\def\sR{{\mathbb{R}}}

\def\sZ{{\mathbb{Z}}}

\DeclareMathOperator*{\argmin}{arg\,min}